\documentclass{article}

\begin{document}

\input epsf.sty



\title{Universal long-time relaxation on lattices 
of classical spins:
Markovian behavior on non-Markovian timescales}

\author{Boris V. Fine\footnote{ 
Department of Physics, University of Illinois at Urbana-Champaign,
Urbana, IL 61801, USA; 
Spinoza Institute, Minnaert Bld., Utrecht University,
Leuvenlaan 4, 3584 CE Utrecht, Netherlands; and 
Max Planck Institute for the Physics of Complex Systems, 
Noethnitzer Str. 38,  D-01187 Dresden, Germany (present address); 
e.mail: fine@mpipks-dresden.mpg.de
}}

\date{September 10, 2002}

\maketitle

\begin{abstract}
The long-time behavior of certain fast-decaying
infinite temperature correlation functions on  one-, two- and
three-dimensional lattices of classical spins with various kinds of
nearest-neighbor interactions is studied numerically, and evidence
is presented that the functional form of this behavior is either
simple exponential or exponential multiplied by cosine. Due to the fast
characteristic timescale of the long-time decay, such a
universality cannot be explained on the basis of conventional
Markovian assumptions. 
It is  suggested that this behavior is related to
the chaotic properties of the spin dynamics. 
\vskip 0.8 cm 
Key words: Spin dynamics, Classical spins, Chaos
\vskip 0.5cm
\end{abstract}

\pagebreak

\section{Introduction}

We perform a numerical study of the long-time behavior of
infinite temperature correlation
functions defined on an infinite lattice of classical spins as:  
\begin{equation} 
G(t)= \langle \;  S_k^x(t) \> \; 
      \hbox{$\sum_{n}$cos({\boldmath $q \cdot r$}$_{kn}$)}
				   \;  S_n^x(0) \; \rangle,
\label{G} 
\end{equation} 
where $S_k^\mu$ is 
the $\mu$th ($x$,$y$ or $z$) spin component on the $k$th lattice site;
{\boldmath $r$}$_{kn}$ is the translation vector between the $k$th and
the $n$th sites;
and {\boldmath $q$} is a wave vector commensurate 
with the lattice periodicity.  We consider three types of 
lattices: a simple 
one-dimensional chain, a two-dimensional square lattice, 
and a three-dimensional cubic lattice.
In each case, the dynamical evolution of the system
is driven by the nearest-neighbor interaction represented by  
the  Hamiltonian
\begin{equation} 
{\cal H} =
\sum_{k,n} [J_x S_k^x S_n^x + J_y S_k^y S_n^y + 
J_z S_k^z S_n^z], 
\label{H} 
\end{equation} 
where $J_\mu$  
are coupling constants. 
With such a Hamiltonian, the timescale of the individual spin motion 
referred to below as the ``mean free time'' can be given by the time
$\tau = [ {1 \over 3} N  S^2 ({J_x}^2+{J_y}^2+{J_z}^2)]^{-1/2}$, 
where $N$ is the number
of the nearest neighbors (twice the number of the lattice dimensions).

In the context of inelastic neutron scattering, correlation
functions (\ref{G}) 
are  called ``intermediate structure factors''\cite{neutron}. 
If $q=0$, Eq.(\ref{G}) can also represent the free induction
decay in nuclear magnetic resonance (NMR)\cite{Abr}.

In this work, we provide extensive 
numerical evidence that the generic long-time behavior of $G(t)$
has one of the following two
functional forms: either
\begin{equation} 
G(t) \simeq e^{- \xi t},
\label{longexp} 
\end{equation} 
or 
\begin{equation} 
G(t) \simeq e^{- \xi t} cos(\eta t + \phi), 
\label{longcos} 
\end{equation} 
where the constants $\xi$ and
$\eta$ are of the order of $1/\tau$.

It is important to realize that, if the above functional form of the
long-time behavior is, indeed, generic (i.e. 
independent
of the specific details of interaction), then
this property is very likely related to the
randomness generated by the spin dynamics. At the same time,
the problem cannot be reduced to
the Markovian paradigm of
``a slow variable interacting with a fast equilibrating
background'' ---  the characteristic timescale $\tau$ in 
Eqs.(\ref{longexp},\ref{longcos})
is not "slow". It is, in fact, the fastest natural  timescale of the problem.
Therefore, whatever is the ultimate explanation of that behavior,
it will certainly be a step 
beyond the standard theory of Brownian-type motion.

Our interest in the long-time behavior of the correlation functions
(\ref{G}) was originally  motivated by two isolated pieces of
evidence supporting 
the oscillatory 
behavior (\ref{longcos}) in {\it quantum} (spin 1/2) systems: 
(i) experiments on NMR free
induction decay in CaF$_2$ \cite{EL} and (ii) the results
of numerical diagonalization of spin 1/2 chains \cite{FLS}.
In the both cases, quantities analogous to the one defined by
Eq.(\ref{G}) have been measured or computed, and the results look
very similar to the plots shown in the left column of Fig.~1.

We came to recognize the importance of a detailed study of the classical
limit, when, in an attempt to explain the long-time relaxation in 
quantum spin systems, we developed a theory that turned
out to be simultaneously applicable to classical spins. 
That theory is  presented in a different paper\cite{Fine2} which 
has been written simultaneously with the present one.
The present paper is mainly numerical: it is 
not intended to be a brief exposition of Ref.\cite{Fine2}.
Below, we only provide the summary
of the results from Ref.\cite{Fine2}.

The theory developed in Ref.\cite{Fine2}
describes long-time relaxation as a correlated diffusion in finite
volumes.  In the classical case, those finite volumes correspond to the
spherical surfaces on which the tips of classical spin vectors 
move, while in the quantum case, the finite volumes originate from a
more sophisticated construction in Hilbert space.  
The overall structure of such a treatment
has noticeable  parallels with the theory of Pollicott-Ruelle resonances 
in classical chaotic systems\cite{Ruelle, Gaspard}.
A definite
prediction from the correlated diffusion description is that the
functional form of the long-time relaxation should be given by
Eqs.(\ref{longexp},\ref{longcos}). 

The important part of the above theory is not the diffusion description
itself but the reason why it is applicable, given the ``non-Markovian''
relaxation timescale.  The theory is based on the fairly strong conjecture
that, for a broad class of many-body systems, a  formal extension of the 
Brownian-like description 
applies to the the long-time
behavior of the {\it ensemble average} quantities, even when
the problem exhibits no separation of the timescales between 
the slow and the fast motions. 

Before proceeding with the description of the simulations, it should be
mentioned that, for the classical spin systems at infinite temperature, 
the long-time behavior of the $q$-dependent
correlation functions (\ref{G}) decaying on the
timescale of $\tau$ has
never been addressed.  
The closest to this subject
was the work of de Alcantara Bonfim and Reiter\cite{BR},
who considered the Heisenberg spin chain and focused
on the long-time behavior of  correlation
functions (\ref{G}) with small $q$. 
Those correlation functions, however,
are not typical  for our purposes, because, as a consequence
of the total spin conservation, they decay
on the timescale, which is much longer than the characteristic 
timescale of one-spin motion. In that situation, 
the hypothesis
of spin diffusion\cite{Bloembergen} would lead 
to the prediction of nearly exponential decay
with the decay constant proportional to $q^2$. 
The results of de Alcantara Bonfim and Reiter did not cover the range
of values, which would be sufficient to confirm or rule out the 
exponential character of the long-time decay. However, those
results (in line with others\cite{Muller},\cite{LG}) indicated that, if, the
spin diffusion regime exists for classical spin chains, the approach
to that regime is anomalously slow.  

The present work
includes one example of the Heisenberg interaction, just to show that
this case does not appear to be special with respect to the long-time
property (\ref{longexp},\ref{longcos}).

\section{Simulations}

Our computational strategy was similar to that of
M\"uller\cite{Muller}. Namely, we did not  deal with very large
systems but, instead, performed an ensemble averaging over a large
number of finite, but not too small, lattices having periodic boundary
conditions.  The finite size effects were then controlled by varying
the size of the lattice.

For a given lattice size, many computational runs have thus been
performed.  Each of them started from completely random initial
conditions (corresponding to the infinite temperature) and generated
the evolution of the system over a time interval two orders of
magnitude longer than the mean free time $\tau$  (see 
Table~1 for specific numbers). The correlation functions were then
obtained by averaging the data within each run and over different runs.

The following algorithm has been used in order to simulate the evolution
of the system.

At each time step, the spins were advanced sequentially 
in such a way that, if the spin number $k$ interacted with 
the spin number
$n$, and the $k$th spin was advanced first, 
then the new coordinates of
the $n$th spin were computed 
based on the local field created by the
already advanced $k$th spin.

The procedure for advancing a given ($k$th)
spin to the next point along the discrete time grid consisted
of two steps.

Step 1: The coordinates of the $k$th spin  were changed by
$\delta${\boldmath$S$}$_k$
according to the straightforward discretization of the equations of
motion, i.e.
\begin{equation}
\delta \hbox{\boldmath$S$}_k = [\hbox{\boldmath$S$}_k \times 
\hbox{\boldmath$h$}_k ] \; \delta t,
\label{timestep}
\end{equation}
where {\boldmath$h$}$_k$ was the local field equal to
$ \sum_n^{\hbox{n.n.}} [ J_x S_n^x \hat{\hbox{\boldmath$e$}}_x +
J_y S_n^y \hat{\hbox{\boldmath$e$}}_y + 
J_z S_n^z \hat{\hbox{\boldmath$e$}}_z ]$ 
(sum over the nearest neighbors of the $k$th spin); 
and $\delta t$ was the discretization time step. 

Step 2: The higher order errors that changed the length of the spin
vector were eliminated.  This was done by contracting the spin
component perpendicular to the local field, so that it took the absolute
value it had before Step 1.

The whole manipulation could not change the spin projection parallel to
the local field and, therefore, the energy of interaction of that spin
with its neighbors.  Since the next spin was advanced in the newly
updated local field, the energy of the whole system was conserved
exactly during the entire integration.  Apart from insuring 
meaningful behavior of the computed trajectories, the exact
conservation of energy substantially improved the convergence of the
algorithm with respect to the limit $\delta t \to 0$.

In our simulations, the discretization time steps (given in Table~1)
were admitted as sufficient when their further reduction
appeared to have no effect on the computed correlation functions.
We also checked that the averaging over a larger number of
much shorter runs led to results consistent with the
longer runs we used.

The time lengths of the runs
indicated in  Table~1 were 
chosen to
optimize the resulting efficiency of averaging:  too short runs (of the
order of the time length of the computed correlation function) 
did not make
many independent contributions to the correlation functions, while 
too long runs did not improve the quality of the
averaging proportionally to their length.

\section{Results}

The results of our simulations for different dimensions, interaction
constants, and wave numbers are presented in Fig.~1\cite{fcc}, together with the
long-time theoretical fits based on either Eq.(\ref{longexp}) or
(\ref{longcos}). 
Since we aimed at demonstrating the exponential
character of the long-time behavior, it was natural to use a
logarithmic scale for $G(t)$. However, because in the half of the
cases, $G(t)$ was also oscillating, we chose to show the logarithmic
plots for the absolute value of $G(t)$, which explains the cusps
in the left column of plots
in Fig.~1. Those cusps correspond to the points where $G(t)$
crosses zero.  The reason that the cusp minima do not 
reach $- \infty$ is that a
discrete grid was used.

The selection of parameters for the simulations was subject to
certain practical constraints:  The computed correlation functions could
be considered reliable only within quite a limited range along both the 
$t$- and Log($G(t)$)- axes. Therefore, we avoided the correlation
functions that reached too-small values too fast, or, on the
contrary, decayed too slowly.

From our experience,
the finite size effects were least pronounced
for correlation functions with $q=0$.
For this reason, $q$ was chosen to be zero in six of eight
examples presented in Fig.~\ref{fig1}.

Since the long-time behavior of the correlation functions was only
marginally accessible with our computational resources, we present the
results in substantial detail, thus making clear the uncertainties
associated with insufficient ensemble averaging and finite size
effects.  Each of the correlation functions presented in
Fig.~\ref{fig1} was computed four times:  two statistically independent
averaging results for each of two different lattice sizes. ``Two
statistically independent averaging results'' means that, in each case,
the same number of sample runs was performed but two different sets 
of random numbers were used for setting the initial
orientations of spins.  Each frame in that figure thus contains a
superposition of four plots.  The time interval where these plots do
not deviate from each other can be considered as representing the limit
of infinite lattice size with sufficient ensemble averaging.

The finite size effects are not
evident in any of the examples shown in Fig.~1 --- 
in every case,
the plots representing different simulation outcomes 
do not deviate from each other before the statistical
fluctuations for each of the two lattice sizes become apparent.

Our experience indicates that further improvement in the accuracy of the
computed correlation functions would simultaneously require a finer
discretization, much more extensive ensemble averaging and, probably,
larger system sizes, i.e. much greater computational effort.

Summarizing the evidence, we observe that, with the marginal exception
of Fig.~1(d), in every other case presented,
there is an interval, covering at least one decade of the values of
$G(t)$, where the simulation results agree with the long-time fits
(\ref{longexp}) or (\ref{longcos}).
We  would also like to point out that, while in all cases the exponential
behavior becomes pronounced quite early, 
the long-time exponents in Figs.~1(e,g) 
describe almost the entire correlation functions.

Thus our
numerical results lend  strong support to
the idea expressed in Ref.\cite{Fine2}, that a discrete spectrum of
well-separated exponents describes the long-time behavior of the
correlation functions considered, with the slowest of those exponents
responsible for the asymptotic functional form given by
Eq.(\ref{longexp}) or (\ref{longcos}).  It is also very likely, though
slightly less reliable, that in the examples presented, our simulations
revealed the slowest exponents. The reservation here is for the
possibility that even slower exponents could enter the long-time
expansion of $G(t)$ with anomalously small coefficients.

\section{Conclusions}

In conclusion, we have presented a numerical evidence that
the long-time behavior of the
correlation functions considered is exponential with or without
the oscillatory component.  Leaving the
details to Ref.\cite{Fine2}, here we just mention that our
best hope  for the theoretical explanation of that
behavior is associated with the strong
chaotic properties of  the spin dynamics. Those properties are likely
to be quite generic, i.e. present in other systems.

\section{Acknowledgements}

The author is grateful to A.~J.~Leggett for numerous discussions,
and to C.~M.~Elliott and R.~Ramazashvili for helpful comments on the manuscript.
This work has been  supported in part by the John~D. and 
Catherine~T. MacArthur Foundation at the University of Illinois,  by 
the Drickamer Endowment Fund at the University of Illinois and by the 
Foundation of Fundamental Research on Matter (FOM), which is sponsored by
the Netherlands Organization for the Advancement of Pure Research (NWO).

\

\textbf{Note added to proof:} After submitting this paper, the author 
became aware of a closely related work of T. Prosen\cite{Prosen}, in which
the decay of the correlation functions defined on 
the lattices of quantum spins
was interpreted in terms of Policott-Ruelle resonances.

\pagebreak

TABLE CAPTIONS:

\

\begin{figure}
\setlength{\unitlength}{0.1cm}
\begin{picture}(70, 140)
{
\put(-20, -40){

\epsfxsize=6.00in
\epsfbox{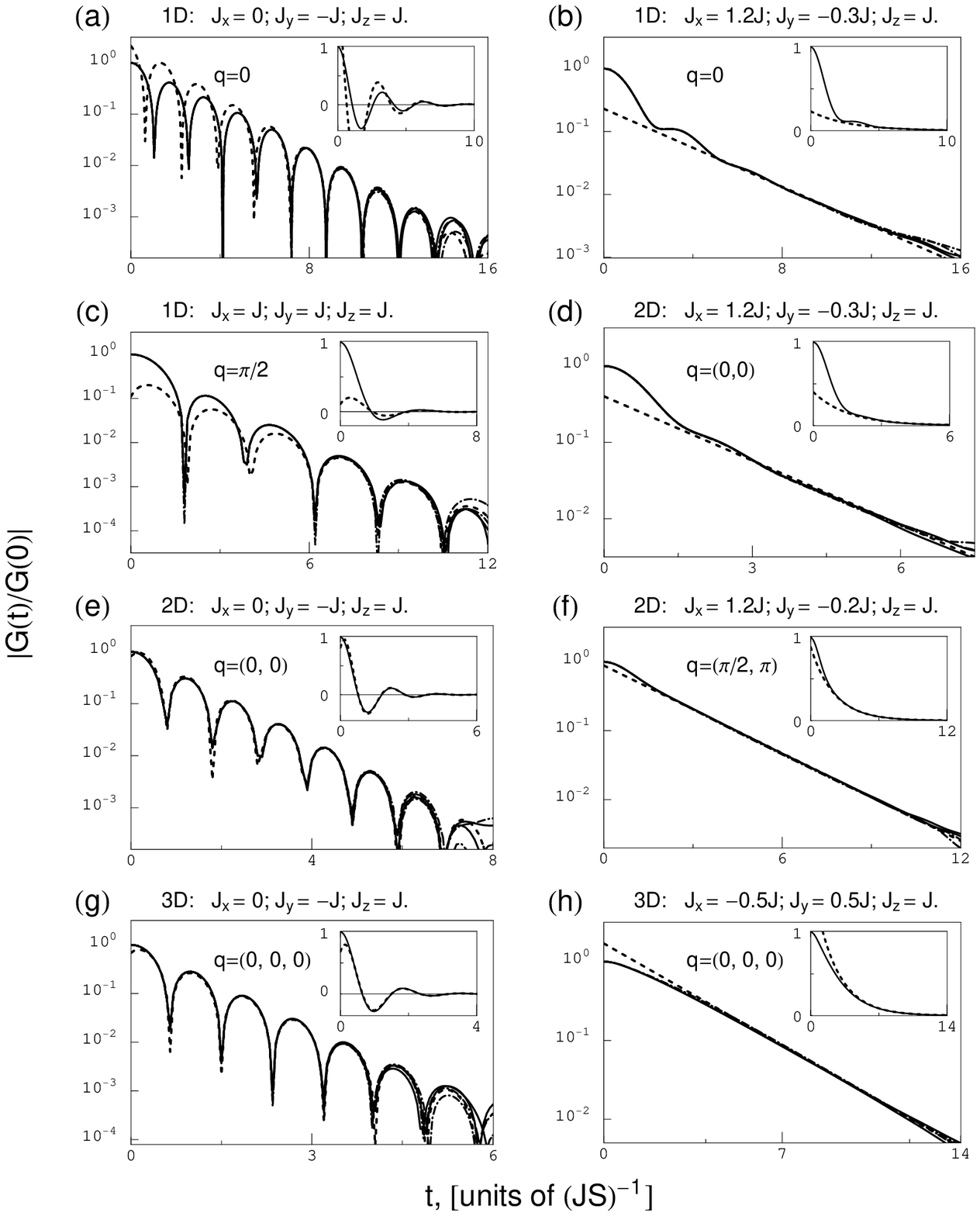} }
}

\end{picture} 
\caption{} 
\label{fig1} 
\end{figure}

Table 1:

Simulation parameters and the long-time fits corresponding to the plots 
presented in Fig.~\protect{\ref{fig1}}. 
The numbers in the right four columns  characterize each of the four
data sets superimposed in the corresponding frame.

\begin{table}
\begin{tabular}{llllll}  
Frame & \multicolumn{2}{c}{Lattice dimensions} & Number & Time length & 
Discreti-  \\ \cline{2-3}
label &  &  & of runs & of each run,  & 
zation   \\
 & Size 1 & Size 2 & & \protect{$[(JS)^{-1}]$} & time step, \\ 
 & & & & & \protect{$[(JS)^{-1}]$} \\ [1mm] \hline \\
(a) & 19 & 15 & 500000 & 320 & 0.02 
\\ [1mm]
(b) & 19 & 15 & 500000 & 320 & 0.02 
\\ [1mm]
(c) & 40 & 24 & 160000 & 320 & 0.02 
\\ [1mm]
(d) & 10 $\times$ 10 & 7 $\times$ 7 & 60000 & 100 & 0.0025 
\\ [1mm]
(e) & 10 $\times$ 10 & 7 $\times$ 7 & 80000 & 170 & 0.01  
\\ [1mm]
(f) & 12 $\times$ 8 & 8 $\times$ 4 & 60000 & 100 & 0.0025  
\\ [1mm]
(g) & 5 $\times$ 5 $\times$ 5 & 4 $\times$ 4 $\times$ 3 & 80000 & 170 & 0.005
\\ [1mm]
(h) & 5 $\times$ 5 $\times$ 5 & 4 $\times$ 4 $\times$ 3 & 80000 & 170 & 0.005
\\ [5mm]
\end{tabular}
\begin{tabular}{ll} 
 \\
Frame &  Long-time fit \\ 
label &  (time units - \protect{$[(JS)^{-1}]$}) \\ [1mm] \hline \\
(a) &  2.25 exp(-0.585 $t$) cos(1.93 $t$ + 0.31)
\\ [1mm]
(b) &  0.23 exp(-0.357 $t$) 
\\ [1mm]
(c) &  0.30 exp(-0.588 $t$) cos(1.46 $t$ - 1.19)
\\ [1mm]
(d) & 0.40 exp(-0.645 $t$) \\ [1mm]
(e) & 1.20 exp(-1.031 $t$) cos(3.06 $t$ - 0.82) \\ [1mm]
(f) & 0.88 exp(-0.488 $t$) \\ [1mm]
(g) & 1.00 exp(-1.299 $t$) cos(3.70 $t$ - 0.82) \\ [1mm]
(h) & 1.70 exp(-0.426 $t$) \\ [1mm]
\end{tabular}
\caption{}
\label{table1}
\end{table}

\pagebreak

FIGURE CAPTIONS:

\

Figure 1:

Correlation functions $G(t)$ of the form (\protect{\ref{G}})
for 1D chain, 2D square lattice, and
3D cubic lattice. The interaction coefficients and the wave
numbers (in the units of inverse lattice spacing) 
are indicated above the plots. The main frame of each figure
shows the logarithmic scale of the absolute value of $G(t)/G(0)$, while
the inset frame shows the direct plots of $G(t)/G(0)$ (with
neither the logarithm nor the absolute value being taken). 
Within each frame, the simulation results
are presented by almost indistinguishable superposition
of data for two lattice sizes, and each size is represented
by two statistically independent averaging results; therefore, two solid lines for the 
larger size (Size 1)
and two dash-dotted lines for the smaller size (Size 2).
The spread of the four lines indicates the
computational uncertainty \protect{---} it becomes visible only in the
lower right corner of each frame.  The dashed lines in each figure are 
the long-time theoretical fits of  form (\protect{\ref{longexp}})
or (\protect{\ref{longcos}}). The numbers relevant to each data set
are given in Table~1.

\pagebreak

\begin{thebibliography}{99}
\bibitem{fcc}
In three dimensions, we have
also computed correlation functions (\ref{G})  for the face-centered  
and the body-centered cubic lattices. The 
nearest-neighbor interaction constants were the same  as those
indicated in Figs.(\ref{fig1}(g,h). 
In every case, we have observed the long-time behavior (\ref{longexp}) or 
(\ref{longcos}).  These results can be obtained from the author.
\end{thebibliography}
\end{document}